\documentclass[english,twocolumn,aps,prb,superscriptaddress,]{revtex4-2}
\usepackage{babel}
\usepackage{amsmath,mathtools}
\usepackage{amssymb}
\usepackage{graphicx}
\usepackage{tabularx}
\usepackage{lipsum}
\usepackage{xcolor}
\begin{document}
\title{Anomalous Non-Hermitian Skin Effects in Coupled Hermitian Chains with Cross-Coupling}
\author{S M Rafi-Ul-Islam }
\email{rafiul.islam@u.nus.edu}
\selectlanguage{english}%
\affiliation{Department of Electrical and Computer Engineering, National University of Singapore, Singapore}
\author{Zhuo Bin Siu}
\email{elesiuz@nus.edu.sg}
\selectlanguage{english}%
\affiliation{Department of Electrical and Computer Engineering, National University of Singapore, Singapore}
\author{Md. Saddam Hossain Razo}
\email{shrazo@u.nus.edu}
\affiliation{Department of Electrical and Computer Engineering, National University of Singapore, Singapore 117583, Republic of Singapore}
\author{Mansoor B.A. Jalil}
\email{elembaj@nus.edu.sg}
\selectlanguage{english}%
\affiliation{Department of Electrical and Computer Engineering, National University of Singapore, Singapore}

\begin{abstract}
In this work, we demonstrate the presence of an anomalous non-Hermitian skin effects which decay from both ends of a system consisting of two coupled Hermitian chains induced by non-reciprocal inter-chain cross-coupling. Another intriguing feature of the system is that its eigenenergy spectrum in thermodynamic limit deviates from the generalized Brillouin zone (GBZ), contrary to that of conventional non-Hermitian systems. The thermodynamic-limit energy spectrum is however restored to the GBZ by the presence of even an infinitesimal amount of gain and/or loss term to the system. In this case, the system exhibits a critical phenomena similar to that of coupled non-Hermitian chains, whereby the eigenspectrum starts to approach the GBZ beyond some critical size. Furthermore, the non Hermitian skin effect becomes less pronounced as the system size exceeds this critical size. We analytically explain these peculiar features which highlight the important role of gain and loss terms as well as inter-chain coupling in tuning non-Hermitian skin modes, thus suggesting a new avenue for the modulation of the skin mode characteristics of open systems.
\end{abstract}
\maketitle

\section{Introduction} 
Much attention has been focused on non-Hermitian systems \cite{ashida2020non,el2018non,PhysRevB.110.045444,gong2018topological} which exhibit various intriguing and novel characteristics such as exceptional points \cite{kawabata2019classification,hu2017exceptional,rafi2021non}, unidirectional transport \cite{longhi2015non,longhi2015robust,du2020controllable}, ultra-sensitive sensing \cite{budich2020non,hokmabadi2019non,yang2021ultrarobust}, amplification of quantum response \cite{wang2022amplification,rafi2024saturation,siegman1989excess}, nodal rings \cite{rafi2021non,carlstrom2018exceptional,wang2019non} and the extensive accumulation of eigenstates at boundaries \cite{okuma2020topological,song2019non,rafi2022interfacial,longhi2020unraveling}, which are absent in their Hermitian counterparts \cite{xu2023photoelectric,rafi2022valley,rafi2023conductance,islam2014thermal,sun2020spin,sun2019field,PhysRevApplied.14.034007}. Among these phenomena associated with non-Hermitian systems, perhaps the most significant one is the last-mentioned one, which is also known as the non-Hermitian skin effect (NHSE). NHSE plays an important roles not only in various aspects of theoretical and experimental studies but also in many potential applications of non-Hermitian systems \cite{yi2020non,deng2022non,rafi2022type,zou2021observation,rafi2022unconventional,yao2018edge}. The NHSE in an open chain is also characterized by eigenenergy spectra which are distinct from those of the corresponding periodic chain and the violation of the conventional bulk-boundary correspondence (BBC) \cite{jin2019bulk,koch2020bulk,kunst2018biorthogonal}. Various aspects of NHSE have been realized in various systems including topolectrical (TE) \cite{helbig2020generalized,rafi2022system,zou2021observation,rafi2024chiral,PhysRevB.107.245114,hofmann2020reciprocal,rafi2024dynamic,rafi2021topological}, photonics \cite{feng2017non,pan2018photonic}, optics \cite{longhi2010optical,parto2021non}, superconducting systems \cite{chen2021quantum,wang2021majorana}, and metamaterials \cite{ghatak2020observation,wen2022unidirectional}. In particular, TE circuits have an advantage over other platforms owing to their accessibility and ease in implementation \cite{rafi2020topoelectrical,lee2020imaging,rafi2020realization,zhang2020topolectrical,rafi2020anti,zhang2023anomalous}.

In recent years, a new type of NHSE with a remarkable finite size dependence, known as the critical NHSE (CNHSE) \cite{li2020critical,rafi2022system,yokomizo2021scaling,liu2020helical,sahin2022unconventional,rafi2022critical}, was realized in coupled non-Hermitian chains. The key feature of the CNSHE is the abrupt (discontinuous) jump in the energy eigenvalue spectrum and the eigenstate spatial localization  when the system size exceeds a critical length. So far, the CNHSE has only been shown to occur in coupled systems in which the individual uncoupled chains both exhibit the NHSE. Observations from these earlier studies led to the conventional belief  that CNHSE arises only if chains with dissimilar NHSE localization lengths are coupled. In this work, we show that this claim does not hold true in general. In addition, our study has uncovered unexpected phenomena - (i) localized non-Hermitian skin modes can be induced in a coupled system consisting of two Hermitian chains which individually do not exhibit NHSE by means of a non-reciprocal inter-chain coupling, and (ii) as abovementioned, a critical phenomena i.e. abrupt jump in the eigenenergy spectrum of the coupled system, is also observed beyond a certain critical size.

\begin{figure*}[htp!]
  \centering
    \includegraphics[width=0.8\textwidth]{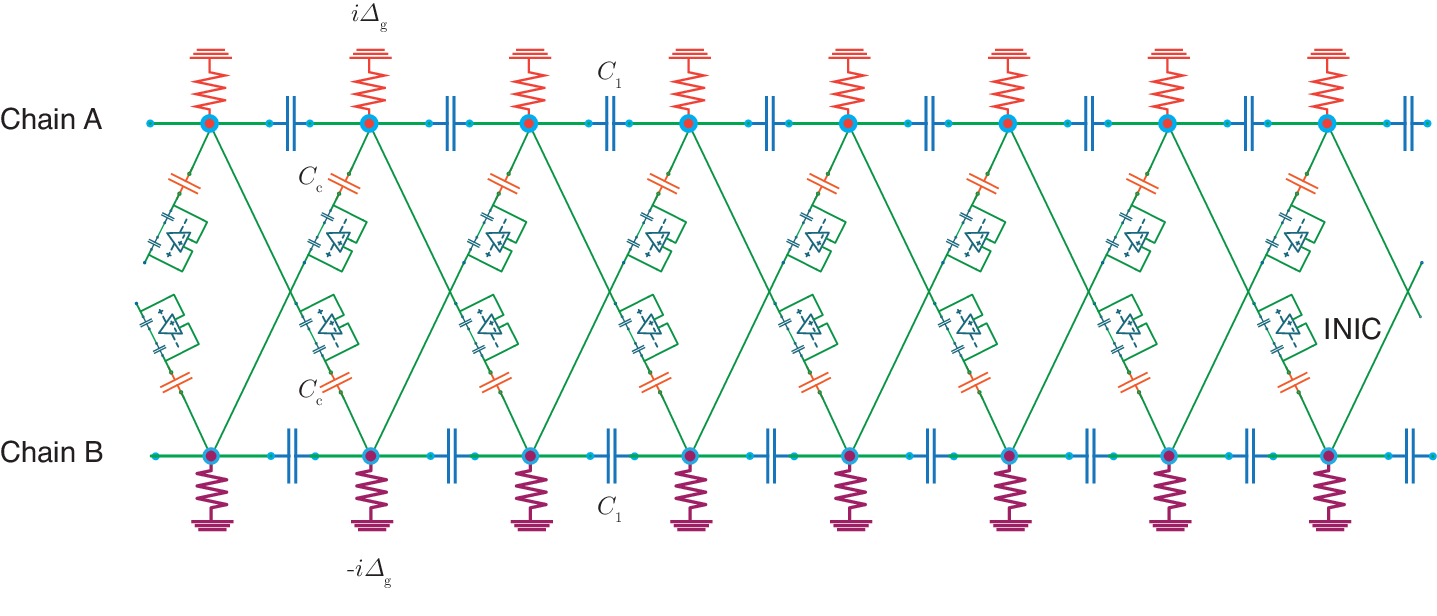}
  \caption{Schematic representation of a pair of cross-coupled 1D chains that exhibits the NHSE. Each individual 1D chain has reciprocal couplings of $C_1$ between each node and its left and right neighbors on the same chain. The INIC realizes a non-reciprocal cross-coupling between each in a chain and its left and right neighbors of its counterpart in the other chain with a coupling magnitude of $C_\mathrm{c}$ and a phase difference of $\pi$ between the coupling of chain A to chain B, and that of chain B to chain A. This non-reciprocal inter-chain cross-coupling results is the key element in inducing the NHSE in the coupled system. In addition, each node on chain A has an on-site loss term of $i\Delta_{\mathrm{g}}$ while each node on chain B has an on-site gain term of $-i\Delta_{\mathrm{g}}$. }
  \label{gFig1}
\end{figure*}  
 
The TE circuit representation of the system under consideration, i.e., a coupled system of two 1D Hermitian chains linked by a cross inter-chain coupling is shown schematically in Fig. \ref{gFig1}. Each node within each of the chains is connnected to its left and right neighbors by a capacitance $C_1$. Each node in the upper (``A'') chain is further coupled to the left and right neighbors of the corresponding node in the lower chain (``B'') by a capacitance $C_{\mathrm{c}}$, whereas each node in chain B is coupled to the left and right neigbors of its counterpart on chain A by a capacitance of equal magnitude but with a phase difference of $\pi$, i.e., $-C_{\mathrm{c}}$. This non-reciprocal coupling between chains A and B can be realized through the use of negative inverters at current inversion (INICs) \cite{rafi2021non}. In addition, each node in chain A (lower) chain has an on-site loss (gain) term of $i\Delta_{\mathrm{g}}$ ($-i\Delta_{\mathrm{g}}$) term which can be realized by grounding with a positive (negative) resistor with a resistance of  magnitude $R_{\mathrm{g}}$ such that $\Delta_{\mathrm{g}}=\frac{1}{\omega R_{\mathrm{g}}}$. 
 
 Remarkably, we find that such a non-reciprocal cross inter-chain coupling can induce non-Hermitian characteristics in the coupled system even when $\Delta_{\mathrm{g}}=0$, i.e., when each individual chain does not only have reciprocal coupling and is hence Hermitian. Although the eigenstates are still localized at the ends of the chains under the non-Hermitian skin effect (NHSE), in contrast to the conventional NHSE in generic one-dimensional non-Hermitian systems, the eigenvalue distribution  of the coupled chains does not coincide with the generalized Brillouin zone (GBZ) in the thermodynamic limit (note that the GBZ gives the loci of complex energies where the middle $|\beta|$ values are the same \cite{yokomizo2023non}.  However, the addition of an even an infinitesimal amount of imaginary on-site gain / loss to the two chains thermodynamic-limit eigenvalue spectrum of the system would restore the thermodynamic-limit eigenspectrum to the GBZ. Thus, in this case, the NHSE weakens as the system size increases and eventually vanishes in the thermodynamic limit.

\section{Results}
The system in Fig. \ref{gFig1} is the analogue of a lattice system with the surrogate Hamiltonian 
\begin{equation}
  H(\beta) = C_1\mathbf{I}_2 \left(\beta + \frac{1}{\beta} \right) + C_{\mathrm{c}} \sigma_x \left(\beta - \frac{1}{\beta} \right) + i \Delta_{\mathrm{g}} \sigma_z \label{Hbeta}
\end{equation}
where $\beta \equiv \exp(i k)$ in which $k$ may, in general, be complex in a non-Hermitian system, $\sigma_x$ is the $x$ Pauli matrix in the chain A--chain B basis, and $\mathbf{I}_2$ is the two-by-two identity matrix. 

\begin{figure*}[htp!]
  \centering
    \includegraphics[width=0.6\textwidth]{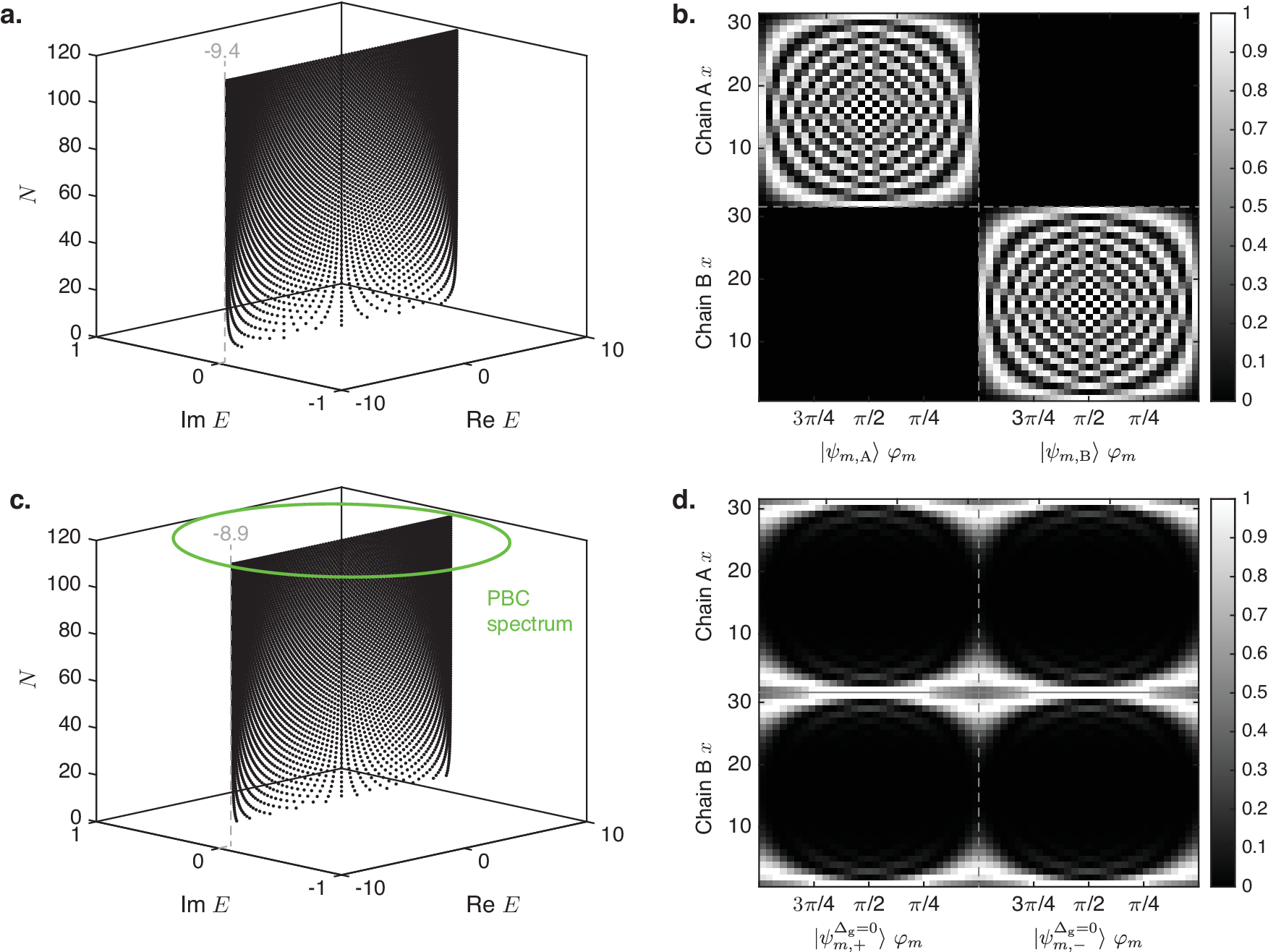}
  \caption{ (a) Eigenspectrum of uncoupled system of two Hermitian chains with $C_{\mathrm{c}}=\Delta_{\mathrm{g}}=0$ and $C_{\mathrm{c}}=1.5$ as a function of the system size $N$. The eigenspectrum lies completely on the real axis. The gray dotted line shows the asymptotic value of the eigenenergy at large $N$. (b) Wavefunction amplitude $|\psi_{\mathrm{A/B}, m}(x)|$. (The wavefunctions are normalized such that the maximum value of $|\psi_{\mathrm{A/B}, m}(x)|$ is 1 for each eigenstate.) (c) Eigenspectrum of two Hermitian chains with $C_{\mathrm{c}}=1.5$ and $\Delta_{\mathrm{g}}=0$ coupled by a non-reciprocal cross-coupling $C_{\mathrm{c}}=1.5$. The eigenspectrum lies completely on the real axis. The PBC spectrum is also shown for comparison. (d) Wavefunction amplitude  $|\psi^{\Delta_{\mathrm{g}}=0}_{m, \pm}(x)|$.   }
  \label{gFig2}
\end{figure*}  

In the absence of the non-reciprocal inter-chain coupling $C_{\mathrm{c}}$ and the gain/loss terms $\pm i \Delta_{\mathrm{g}}$, the system reduces to two uncoupled chains. Under open boundary conditions (OBC) where each chain has a finite length of $N$ sites, Eq. \eqref{Hbeta} corresponds to the lattice Hamiltonian for two uncoupled copies of the 1D free electron gas in an infinite potential well. In such an infinite potential well, the allowed values of the momentum $k$ are quantized to the real values of $k_m = m \pi/(N+1)$, where $m$ is an integer from 1 to $N$ with the corresponding eigenvalues of $2C_1 \cos(k_n)$ (Fig. \ref{gFig2}a). The corresponding eigenstate wavefunctions on chain A / B takes the form of $\psi_{m, \mathrm{A,B}}(x) = \sin(k_m x)$ where $x$ ranges from $1$ to $N$. In such a system, the wavefunction is delocalized across the bulk of the system because the real values of $k$ implies an absence of exponential localization. In other words,  there is no NHSE since the entire system is Hermitian (Fig. \ref{gFig2}b).

We now examine how the eigenspectrum of the system is modified when the non-Hermitian inter-chain cross-coupling is turned on by setting $C_{\mathrm{c}}$ to a finite value while still retaining $\Delta_{g}=0$ . To do this, we first note that $H(\beta)$ in Eq. \eqref{Hbeta} satisfies 
\begin{equation}
  H(\beta)=\sigma_z H(\beta^{-1})\sigma_z. \label{szHsz}
\end{equation}
Because $\sigma_z$ is unitary and is its own inverse, Eq. \eqref{szHsz} implies that if $H(\beta)$ has an eigenvector $|\chi\rangle$ with an eigenvalue $E$, then $H(\beta^{-1})$ also has the same eigenvalue $E$ with the eigenvector $\sigma_z|\chi\rangle$. We shall show that the system undergoes a critical transition as $\Delta_{\mathrm{g}}$ is switched from 0 to an infinitesimally small value. We consider the $\Delta_{\mathrm{g}}=0$ case first. 

When $\Delta_{\mathrm{g}}=0$, the eigenvalues $E_\pm$ of $H(\beta)$ and their corresponding eigenvectors $|\chi_\pm\rangle$  are given by 
\begin{align}
  E^{\Delta_{\mathrm{g}}=0}_\pm &= C_1 \left( \beta + \frac{1}{\beta }\right) \pm C_{\mathrm{c}} \left( \beta - \frac{1}{\beta}\right) \label{EDg0},\\
  |\chi^{\Delta_{\mathrm{g}}=0}_\pm\rangle &= \begin{pmatrix} \pm 1 \\ 1 \end{pmatrix} \label{chiDg0}.
\end{align}

Solving Eq. \eqref{EDg0} for a given value of $E$ results in a fourth-order polynomial in $\beta$ with the solutions 
\begin{equation}
  \beta^{\Delta_{\mathrm{g}}=0}_{s_a,s_b} = \frac{E + s_a \sqrt{E^2 + 4 (C_{\mathrm{c}}^2-C_1^2)  }}{2(C_1 + s_b C_{\mathrm{c}})} \label{betaDg0}
\end{equation}
where $s_a$ and $s_b$ can independently take the values of $\pm 1$, thus yielding the four solutions of $\beta$. Notice that when the terms in the square root of Eq. \eqref{betaDg0} is negative and $E$ is real, i.e., when $E^2 < 4(C_1^2-C_{\mathrm{c}}^2)$, the $\beta^{\Delta_{\mathrm{g}}=0}_{\pm 1, s_b}$s for a given $s_b$ form a complex conjugate pair with

\begin{align}
  \left|\beta^{\Delta_{\mathrm{g}}=0}_{s_a, s_b}\right| &= \left |\frac{C_1-C_{\mathrm{c}}}{C_1+C_{\mathrm{c}}}\right |^{\frac{s_b}{2}} \label{absBetaDg0} \\
  \mathrm{arg}\left( \beta^{\Delta_{\mathrm{g}}=0}_{s_a, s_b} \right) &= s_a \varphi,\ \varphi \equiv \cos^{-1} \left( \frac{2E}{ \sqrt{C_1^2-C_{\mathrm{c}}^2}} \right). \label{argBetaDg0}
\end{align}

Eqs. \eqref{absBetaDg0} and \eqref{argBetaDg0} imply that $\beta^{\Delta_{\mathrm{g}}=0}_{s_a, s_b}$ is the inverse of $\beta^{\Delta_{\mathrm{g}}=0}_{-s_a, -s_b}$. Following the arguments after Eq. \eqref{szHsz}, this implies that $\beta_{s_a,s_b}$ and $\beta_{-s_a,s_b}$ for a given pair of $(s_a, s_b)$ at a given $E$ have the eigenspinor $(1,1)^{\mathrm{T}}$ while $\beta_{s_a,-s_b}$ and $\beta_{-s_a,-s_b}$ are associated with $(1,-1)^{\mathrm{T}}$ - specifically, $\beta_{s_a,s_b}$ is associated with $(1, s_b)^{\mathrm{T}}$. Accordingly, in an OBC system with $N$ nodes in each chain at $x=1$ to $N$, the eigenstates are two-fold degenerate and can be labelled by $s_b$ and the quantum number $m$, i.e., $|\psi^{\Delta_{\mathrm{g}}=0}_{m, s_b}\rangle$ ($m \in (1,..,N)$). The corresponding wavefunction $\psi^{\Delta_{\mathrm{g}}=0}_{m,s_b}(x)$ satisfies the boundary conditions $\psi^{\Delta_{\mathrm{g}}=0}_{m, s_b} (0) = \psi^{\Delta_{\mathrm{g}}=0}_{s_b} (N+1) = (0,0)^{\mathrm{T}}$ and are given by  
\begin{equation}
  \psi^{\Delta_{\mathrm{g}}=0}_{m,s_b}(x) = |\beta^{\Delta_{\mathrm{g}}=0}_{+1,s_b}|^x \sin(\varphi_m x) \begin{pmatrix} 1 \\ s_b \end{pmatrix} \label{psiDg0}
\end{equation}
where $\varphi_m = m\pi/(N+1)$, and 
\begin{equation}
  E_m = 2\sqrt{C_1^2-C_{\mathrm{c}}^2}\cos(\varphi_m). \label{EmDg0}
\end{equation}
Similar to the uncoupled case considered earlier,  the energy eigenvalue distribution remains on the real axis although they are now bounded within a narrower range of $\pm 2\sqrt{C_1^2-C_{\mathrm{c}}^2}$ regardless of the system size (Fig. \ref{gFig2}c).

A significant change is induced by the introduction of the non-reciprocal inter-chain coupling. Even though the individual uncoupled chains are Hermitian, the system now exhibits the NHSE in which the wavefunctions of the eigenstates are localized near the ends of the two chain with an inverse decay length of $1/|\beta^{\Delta_{\mathrm{g}}=0}_{+1,s_b}|$, independent of the system size. In particular, because $|\beta^{\Delta_{\mathrm{g}}=0}_{+1,+1}| = |\beta^{\Delta_{\mathrm{g}}=0}_{+1,-1}|^{-1}$, the states associated with the larger value of $|\beta^{\Delta_{\mathrm{g}}=0}_{+1,s_b}|$ for $s_b \in (+1,-1)$ are localized near the right edge of the system, while those associated with the smaller value of $|\beta^{\Delta_{\mathrm{g}}=0}_{+1,s_b}|$ are localized near the left edge of the system. Note that because $|\psi^{\Delta_{\mathrm{g}}=0}_{m, +1}\rangle$ has the same eigenenergy as  $|\psi^{\Delta_{\mathrm{g}}=0}_{m, -1}\rangle$, any linear combination of $|\psi^{\Delta_{\mathrm{g}}=0}_{m, +1}\rangle$ and $|\psi^{\Delta_{\mathrm{g}}=0}_{m, -1}\rangle$ is also an eigenstate with the same eigenenergy. In particular, 
\begin{widetext}
\begin{equation}
    \psi^{\Delta_{\mathrm{g}}=0}_{m,\pm}(x) =  \sin(\varphi_m x) \left( \left|\beta^{\Delta_{\mathrm{g}}=0}_{+1,-1}\right|^{N/2}\left|\beta^{\Delta_{\mathrm{g}}=0}_{+1,+1}\right|^x   \begin{pmatrix} 1 \\ 1 \end{pmatrix} + \left|\beta^{\Delta_{\mathrm{g}}=0}_{+1,+1}\right|^{-N/2}\left|\beta^{\Delta_{\mathrm{g}}=0}_{+1,-1}\right|^x \begin{pmatrix} 1 \\ -1 \end{pmatrix}  \right) \label{psiDg0pm}
\end{equation}
\end{widetext}
are eigenstates of the system. Noting that Eqs. \eqref{absBetaDg0} and \eqref{argBetaDg0} together imply that $|\beta^{\Delta_{\mathrm{g}}=0}_{+1,-1}|$ in the second term of Eq. \eqref{psiDg0pm} is the inverse of  $|\beta^{\Delta_{\mathrm{g}}=0}_{+1,+1}|$ in the first term when $E^2 < 4(C_1^2-C_{\mathrm{c}}^2)$, Eq. \eqref{psiDg0pm} represents a wavefunction in which the eigenstates symmetrically decay exponentially from \textit{both} ends of the system with the two NHSE localization decay lengths of $\pm \log |\beta^{\Delta_{\mathrm{g}}=0}_{+1,+1}|$  (Fig. \ref{gFig2}d). This is in marked contrast from an eigenstate of a generic 1D non-Hermitian system  where the eigenenergy spectrum in the thermodynamic limit is given by the loci of complex energy plane where the middle $|\beta|$ values coincide and there is a only a single NHSE inverse decay length given by the logarithm of the common middle value of $\beta$, which leads to the eigenstate being localized at only \textit{one} edge of the system. However, it should be noted that not all forms of anti-reciprocal inter-chain coupling between otherwise Hermitian chains will inherently lead to the non-Hermitian skin effect (NHSE). We provide a counter-example in the Appendix where we show that a direct inter-chain coupling between two Hermitian chains where the $j$th node is coupled to its counterpart $j$th node on the B chain does not induce NHSE in the coupled system.

Another significant fact about this system is that the energy eigenvalues do not coincide with the GBZ of the system, even in the thermodynamic limit. To show this, we first label the four $\beta$ values as $\beta_i,\ i\in(1,...,4)$ with $|\beta_1|\leq|\beta_2|\leq ... |\beta_4|$. The  GBZ of the system is then defined as the set of states at which $|\beta_2|=|\beta_3|$. In our system where $H(\beta)$ and $H(1/\beta)$ have the same eigenvalues, we have $|\beta_1| = 1/|\beta_4|$ and $|\beta_2|=1/|\beta_3|$. By definition, $|\beta_2|=|\beta_3|$ on the GBZ  while the equality $|\beta_2|=1/|\beta_3|$ in our system implies that the complex energy plane GBZ of the system is given by the locus of energies at which $|\beta_2|=|\beta_3|=1$. This corresponds to real values of $k$, i.e., $k=-i\ln(\beta)$ and hence, the periodic boundary condition (PBC) spectrum of the system. Putting $\beta=\exp(ik)$ into Eq. \eqref{Hbeta} and diagonalizing gives the PBC spectrum as 
\begin{equation}
  E^{\mathrm{PBC}}(k) = 2 C_1\cos(k) \pm i \sqrt{ (2 C_{\mathrm{c}} \sin(k))^2 + \Delta_{\mathrm{g}}^2 }, \label{Epbc}
\end{equation}
which takes the form of an ellipse with the diameters of $2C_1$ along the real energy axis and $2C_{\mathrm{c}}$ along the imaginary axis when $\Delta_{\mathrm{g}}=0$. This differs markedly from the eigenenergy locus of a line along the real energy axis bounded between $\pm \sqrt{C_1^2-C_{\mathrm{c}}^2}$ in Eq. \eqref{EmDg0}. As a side remark, we note that Eq. \eqref{Epbc}, cannot be the eigenenergy spectrum when $\Delta_{\mathrm{g}}=0$. This is because $\beta_1$ and $\beta_2$ share the same eigenspinor, say $(1, s_b')^{\mathrm{T}}$, while $\beta_3$ and $\beta_4$ share the eigenspinor $(1, -s_b')^{\mathrm{T}}$, which is linearly independent from that of $\beta_1$ and $\beta_2$. This linear independence implies that a wavefunction containing finite weights of all four $\beta$ values, $\psi(x) = (a_1 \beta_1^x + a_2 \beta_2^x) (1, s_b')^{\mathrm{T}} + (a_3 \beta_3^x + a_4 \beta_4^x) (1, -s_b')^{\mathrm{T}}$ with $|\beta_i|>0 \forall i \in (1,2,3,4)$ can never cancel off to zero to satisfy the OBC of $\psi(0) = \psi(N+1) = (0,0)^{\mathrm{T}}$ immediately beyond the extent of the chain except when $a_1$ and $a_2$ are finite and $a_3=a_4=0$,  or vice-versa, as in Eq. \eqref{psiDg0}, or as a linear combination of the wavefunctions, as in Eq. \eqref{psiDg0pm}. 

\begin{figure*}[htp!]
  \centering
    \includegraphics[width=0.8\textwidth]{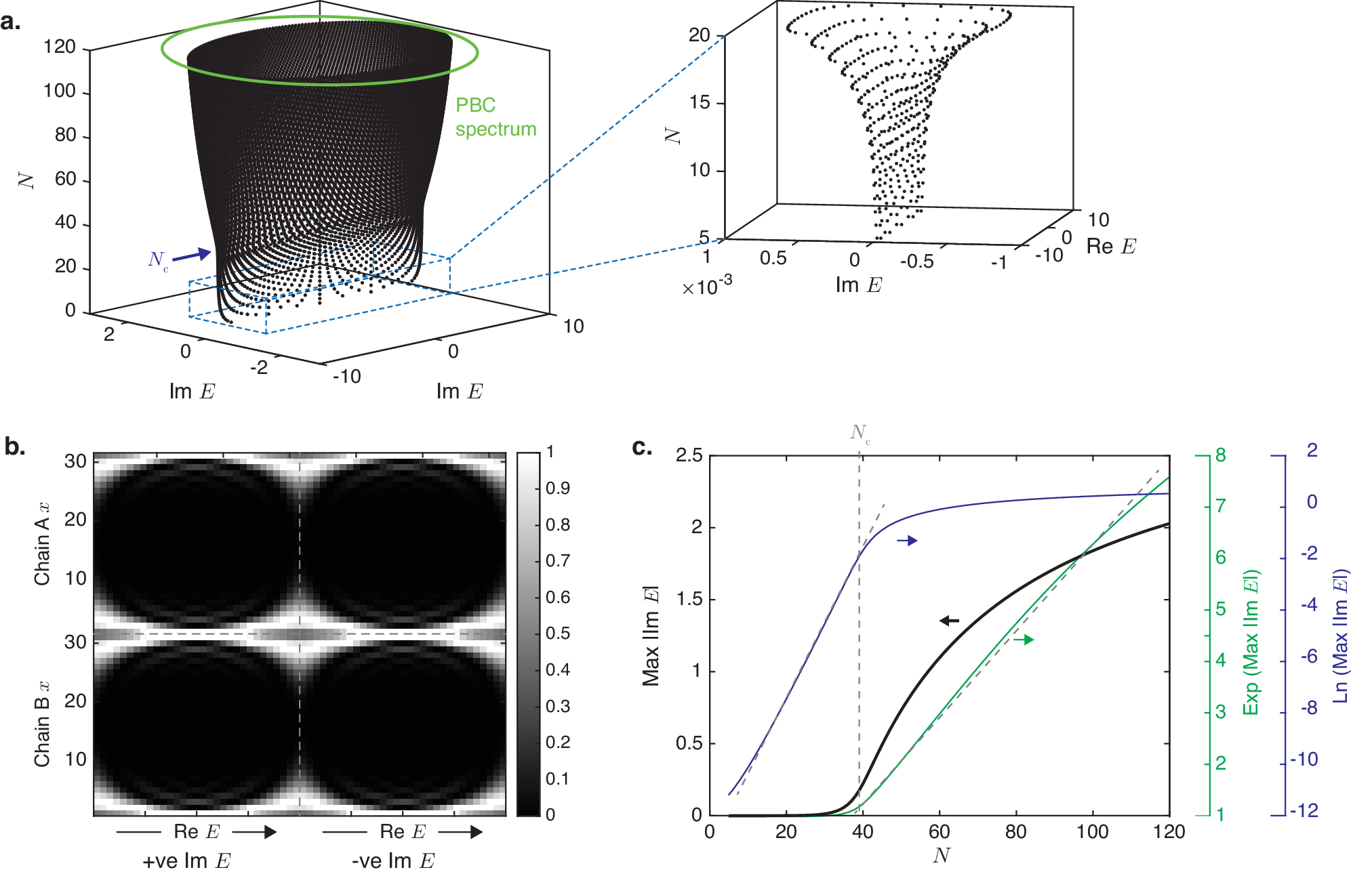}
  \caption{ (a) Eigenspectrum of two reciprocal chains with $\Delta_{\mathrm{g}}=10^{-5}$, $C_{\mathrm{1}}=1.5$ coupled by a non-reciprocal cross-coupling of $C_{\mathrm{c}}=1.5$ as a function of the system size $N$. The PBC eigenenergy spectrum is also shown for comparison, and the critical system size $N_{\mathrm{c}}$ indicated. The inset shows that there are small finite imaginary parts in the eigenenergy. (b) Eigenstate wavefunction amplitude of system in a.  (c) The maximum moduli of the imaginary parts of the eigenenergies $\mathrm{Max}|\mathrm{Im}(E)|$, and its logarithm and exponent for the system in a. The dotted lines serve as guides to the eye.  }
  \label{gFig3}
\end{figure*}

The introduction of an infinitesimal amount of $\Delta_{\mathrm{g}}$ breaks the pair-wise equality of the eigenvectors of $H(\beta)$, which results in the thermodynamic-limit eigenenergy spectrum of the system reverting to its GBZ. To show this in detail, consider  the eigenspinors in the presence of a finite $\Delta_{\mathrm{g}}$ and $C_{\mathrm{c}}$, which  to linear order in $\Delta_{\mathrm{g}}$ is given by 
\begin{equation}
  |\chi_{s_a,s_b}\rangle = \begin{pmatrix} 1 \\ s_b \end{pmatrix} + is_b\Delta_g \begin{pmatrix} 0 \\ \frac{C_1^2-C_{\mathrm{c}}^2}{C_{\mathrm{c}}(C_{\mathrm{c}} E - s_a C_1 \sqrt{4 C_{\mathrm{c}}^2 - 4C_1^2 + E^2})} \end{pmatrix}. \label{chiDg}
\end{equation}

Eq. \eqref{chiDg} shows that the introduction of even an infinitesimal finite amount of $\Delta_{\mathrm{g}}$ causes the eigenspinors for all four possible combinations of $s_a$ and $s_b$ to differ from one another. This implies that in contrast to the $\Delta_{\mathrm{g}}=0$ eigenstates in Eq. \eqref{psiDg0}, which contain only the two bulk eigenstates sharing the same eigenspinor, a finite-$\Delta_{\mathrm{g}}$ eigenstate would require all four $\beta$ values to have finite weights in order to satisfy the boundary conditions $\psi(N+1)=\psi(0) = (0,0)^{\mathrm{T}}$ at both ends of the system. 

We find numerically that at small values of $N$, the locus of the eigenenergies on the complex energy plane takes a similar form to that of the $\Delta_{\mathrm{g}}=0$ case comprising a set of points lying close to the real energy plane (Fig. \ref{gFig3}a). In contrast to the $\Delta_{\mathrm{g}}=0$ case where the eigenenergies lie exactly on the real energy axis and are two-fold degenerate because the two chains are identical to each other,  in the finite-$\Delta_{\mathrm{G}}$, the energy degeneracy is broken  due to the $2i\Delta_{\mathrm{g}}$ on-site potential difference between the two chains. The lifting of the degeneracy  causes the eigenvalues of the latter to be symmetrically distributed about both sides the real energy axis with finite imaginary components (see magnified inset of Fig. \ref{gFig3}a). Interestingly, in the coupled system, the presence of onsite dissipation modifies the solution of the non-Bloch factor  $\beta$. Consequently, the inverse decay length of the NHSE varies  as a function of  $\Delta_{\mathrm{G}}$ and can be expressed as:

\begin{equation}
\begin{aligned}
|\beta| &= e^{-i \xi_{onsite}}  \\ 
        &  = \sqrt{\left|\frac{2 C_1^2+2C_c^2+ \Delta_g^2 + \sqrt{(4C_c^2+\Delta_g^2)(\Delta_g^2+4C_1^2)}}{2(C_1^2-C_c^2)}\right|}
\label{oneq3}
\end{aligned}
\end{equation}
where,  $\xi_{onsite}$ denotes the inverse decay length of the eigenstates localization. Eq. \ref{oneq3} reveals a remarkable modification of the inverse decay length of the skin modes, as the gain and loss terms vary.  Interestingly, in the decoupled limit,  $\xi_{onsite}$ simplifies to  $\xi_{onsite}=-\frac{1}{2} \log \left|\frac{\sqrt{4C_1^2-\Delta_g^2}+i \Delta_g}{2 C_1} \right|=0$. Therefore, onsite potentials does not induce the localization of non-Hermitian skin modes  in the decoupled limit (see Appnedix: B for details).   
 
In the coupled system, the eigenenergies have  slightly larger magnitude in chain A (B) with a positive (negative) imaginary on-site potential for those eigenenergies with positive (negative) imaginary parts. The similarity of the energy spectra and the wavefunctions between the finite $\Delta_{\mathrm{g}}$ case at small sizes (Fig. \ref{gFig3}a, b) and the $\Delta_{\mathrm{g}}=0$ case (Fig. \ref{gFig2}c, d) suggests that the eigenstates of the former can be understood as slightly perturbed versions of the $\Delta_{\mathrm{g}}=0$ eigenstates, as we will explain in more detail later. The magnitude of the imaginary component of the eigenvalue $E$ increases exponentially with the system size $N$ until a critical size $N_{\mathrm{c}}$, as evidenced by the linearity of the $\mathrm{Ln}\ \mathrm{Max}\ |\mathrm{Im}\ E|$ curve for $N < N_{\mathrm{c}}$ in Fig. \ref{gFig3}c. ($\mathrm{Max}\ |\mathrm{Im}\ E|$ occurs at the eigenstates with purely imaginary energies, as can be seen from Fig. \ref{gFig3}a.) 

 When the system size is further increased beyond $N_{\mathrm{c}}$, the eigenenergies in the finite $\Delta_{\mathrm{g}}$ case begin to exhibit larger deviations from those in the $\Delta_{\mathrm{g}}=0$ case. The eigenenergies have progressively larger imaginary components, and begin to approach the complex energy plane GBZ, which coincides with the ellipse of the PBC spectrum (Fig. \ref{gFig3}a). The $\exp(\mathrm{Max}\ |\mathrm{Im}\ E|)$ curve in Fig. \ref{gFig3}c shows that the magnitude of the imaginary components now increases in an approximately logarithmic  rather than exponential manner with respect to  the system size.  This size-dependent switchover in the behavior of the eigenenergy spectrum at a critical system size (i.e. at $N = N_c$) is reminiscent of the switchover phenomena previously observed in two other related non-Hermitian systems. In the critical non-Hermitian skin effect (CNHSE) \cite{li2020critical,rafi2022critical}, the energy spectrum of a system comprising two non-reciprocal 1D chains with different GBZs which are coupled in parallel (i.e., each node on one chain is coupled to its counterpart on the other chain) is similar to the union of the GBZs of the uncoupled chains below the critical system size but abruptly begins to approach the GBZ of the coupled system above the critical size. In the terminal-coupling-induced energy spectrum transition (TCIEST) \cite{siu2023terminal}, the energy spectrum of a single 1D non-reciprocal chain with its two ends weakly coupled  to form a closed loop, exhibits a switchover from the OBC spectrum  to the PBC one when the system size exceeds the critical size. In both the CNHSE and the TCIEST, the changeover from the small- to the large-size regime spectra can be attributed to the exponential localization due to the NHSE. This exponential localization results in an increasing difference in the wavefunction magnitudes of the coupled sites (the end nodes of one chain and their counterparts on the other chain in the CNHSE, and the two ends of the chain in the TCIEST) as the system size increases. At the critical system size, this difference becomes overwhelmingly large for the boundary conditions to be satisfied and the small-size regime energy spectrum can no longer be sustained. The system abruptly shifts to the large-size regime, where the eigenenergies now gain imaginary parts, which increase with the system size, so as to allow  an approximately constant magnitude difference in the wavefunction to be maintained between the coupled sites regardless of the system size. As we shall show analytically in detail, the transition between the small- and large-size regime in the system  in the present work can also be related to the exponential localization due to the NHSE. However, the system studied in the present work differs from those in the CNHSE and TCIEST in one key aspect - the 1D chains in the CNHSE and TCIEST have non-reciprocal couplings, while in the present system the intra-chain couplings within the individual chains are reciprocal and non-reciprocity only occurs in  the inter-chain coupling. 

 The analytical explanation of  the aforementioned transition from the small-size regime  to the large-size regime is as follows. The small-size regime is characterized by a similar eigenenergy distribution over the complex energy plane, and a similar spatial wavefunction distribution as that of the $\Delta_{\mathrm{g}}=0$ case, while in the large-size regime the eigenenergy distribution approaches the PBC spectrum. For convenience, let us shift the $x$ axis so that the chain now lies between $x = -(N-1)/2$ to $(N-1)/2$ instead of $x=1$ to $N$ for convenience. The  wavefunction of an eigenstate in the $\Delta_g=0$ case in Eq. \eqref{psiDg0pm} can then be written as 
\begin{widetext}
 \begin{align}
    & \psi^{\Delta_{\mathrm{g}}=0}(x) \nonumber \\
		&= \left|\beta^{\Delta_{\mathrm{g}}=0}_1\right|^x \left( c_1\exp(i \mathrm{Re}(k_1) x) \begin{pmatrix}1 \\ 1 \end{pmatrix} + c_2 \exp(i \mathrm{Re}(k_2) x) \begin{pmatrix}1 \\ 1 \end{pmatrix} \right) \nonumber \\
    &+   \left|\beta^{\Delta_{\mathrm{g}}=0}_1\right|^{-x} \left( c_3 \exp(-i \mathrm{Re}(k_1) x) \begin{pmatrix}1 \\ -1 \end{pmatrix} + c_4 \exp(-i \mathrm{Re}(k_2) x) \begin{pmatrix}1 \\ -1 \end{pmatrix} \right) \label{psiXsymNDg0} 
 \end{align}
\end{widetext}
where $|\beta_1|\leq|\beta_2|...\leq|\beta_4|$. Here,  we assumed that $\beta_1$ corresponds to the eigenspinor $(1,1)^\mathrm{T}$ and made use of the fact that $\beta_4=1/\beta_1$, $\beta_3 = 1/\beta_4$. In particular, when $\Delta_{\mathrm{g}}=0$,  $|\beta^{\Delta_{\mathrm{g}}=0}_1|=|\beta^{\Delta_{\mathrm{g}}=0}_2|$, and $|\beta^{\Delta_{\mathrm{g}}=0}_3|=|\beta^{\Delta_{\mathrm{g}}=0}_4|$ . The introduction of a finite $\Delta_g$ breaks the equality between $|\beta_1|$ and $|\beta_2|$, and that between $|\beta_3|$ and $|\beta_4|$.  $|\beta_1|$ and $|\beta2|$ can then be written as $|\beta_1| = |\bar{\beta}_{1,2}|/|r_{1,2}|$ and $|\beta_2| = |\bar{\beta}_{1,2}|r_{1,2}$, where $|\bar{\beta}_{1,2}|\equiv \sqrt{|\beta_1||\beta_2|}$ and $r_{1,2} = \sqrt{|\beta_2|/|\beta_1|}$, and analogously for $|\beta_{3,4}|$. (We omit the explicit expressions for $|\bar{\beta}_{1,2}|$ and $r_{1,2}$ because they are cumbersome and not particularly illuminating. To linear order in $\Delta_{\mathrm{g}}$, $|\bar{\beta}_{1,2}| = |\beta_1^{\Delta_{\mathrm{g}}=0}|$.) As shown earlier in Eq. \eqref{chiDg}, the finite $\Delta_{\mathrm{g}}$ also results in a deviation of the eigenspinors from $(1,\pm 1)^\mathrm{T}$. We write the eigenspinors corresponding to $\beta_1$ to $\beta_4$ as $(1, X_i)^\mathrm{T}$ where the relation  $H(\beta)=\sigma_zH(1/\beta)\sigma_z$ implies that $X_3=-X_2$ and $X_4=-X_1$. A finite $\Delta_{\mathrm{g}}$ therefore modifies Eq. \eqref{psiXsymNDg0} to 
\begin{widetext}
 \begin{align}
    & \psi(x) \nonumber \\
		&= \left|\bar{\beta}_{1,2}\right|^x \left( c_1 r_{1,2}^{-x} \exp(i \mathrm{Re}(k_1) x_1) \begin{pmatrix}1 \\ X_1 \end{pmatrix} + c_2 r_{1,2}^{x} \exp(i \mathrm{Re}(k_2) x) \begin{pmatrix}1 \\ X_2 \end{pmatrix} \right) \nonumber \\
    &+  \left|\bar{\beta}_{1,2}\right|^{-x} \left( c_3 r_{1,2}^{-x} \exp(-i \mathrm{Re}(k_1) x) \begin{pmatrix}1 \\ -X_2 \end{pmatrix} + c_4  r_{1,2}^{x} \exp(-i \mathrm{Re}(k_2) x) \begin{pmatrix}1 \\ -X_1 \end{pmatrix} \right) \label{psiXsymNDg0a}. 
 \end{align}
\end{widetext}

\begin{figure*}[htp!]
  \centering
    \includegraphics[width=0.9\textwidth]{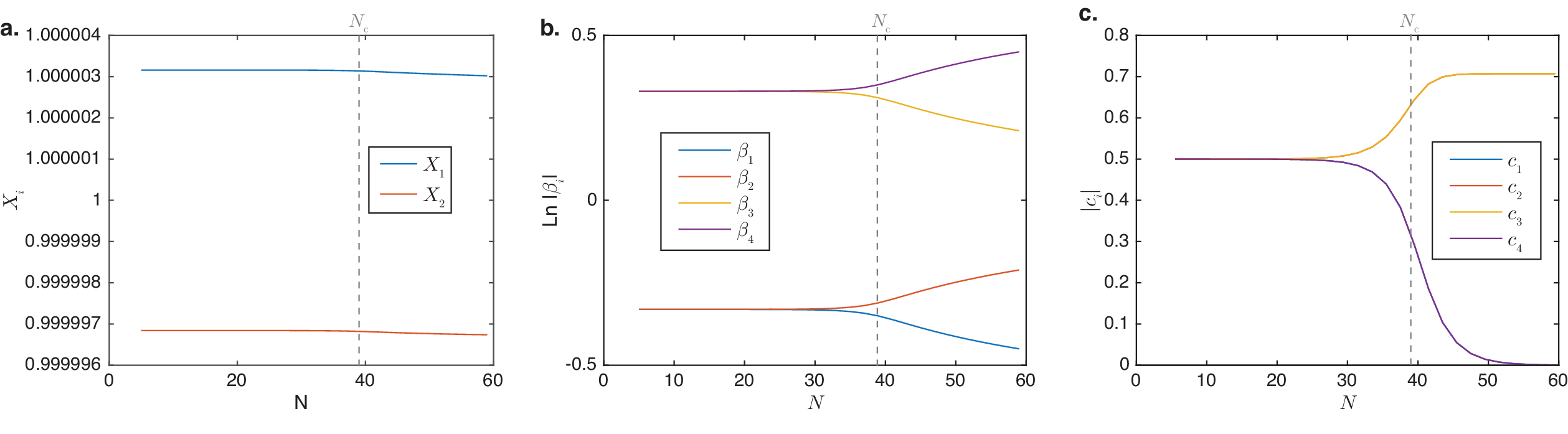}
  \caption{ (a) $X_1$ and $X_2$ ($X_3=-X_2$, $X_4=-X_1$), (b) $\mathrm{Ln}\ |\beta_i|$s, and (c) $|c_i|$s plotted against $N$ for the eigenstates of the system in Fig. \ref{gFig3} with positive purely imaginary eigenenergies. (In (c). the $|c_1|$ curve overlaps with the $|c_4|$ curve, while the $|c_2|$ curve overlaps with the $|c_3|$ curve)  The grey vertical dotted line indicates the value of $N_{\mathrm{c}}$. }
  \label{gFigWfc}
\end{figure*}  

Fig. \ref{gFigWfc} shows the numerically computed values of the $X_i$s, $c_i$s, and $\beta_i$s for the eigenstates with positive purely imaginary eigenenergies for the system shown in Fig. \ref{gFig3}. In Fig. \ref{gFigWfc}a, the values of the $X_1$ and $X_2$ deviate from 1 in the $\Delta_{\mathrm{g}}=0$ case by a very small amount. The rate of variation with the system size $N$ is also very slow, although there is a slight increase when $N > N_c$. Thus, the transition of the system from the small- to large-size regime is not primarily driven by the energy dependence of the eigenspinors. Fig. \ref{gFigWfc}b shows that for system sizes in the small-size regime, i.e.  well below $N_{\mathrm{c}}$, the values of $|\beta_i|$ vary very slowly with energy and $|\beta_1|$ ($|\beta_3$) deviates from $|\beta_2|$ ($|\beta_4$) only minimally. Similarly,  Fig. \ref{gFigWfc}c shows that at small system sizes, all the four $c_i$s have approximately the same magnitude. All the features observed at $N < N_c$, i.e., the  small deviations of $X_{1,2}$ from 1 and the small disparities between $|\beta_1|$ and $|\beta_2|$, between $|\beta_3|$ and $|\beta_4|$, and between the four values of $|c_i|$s, account for the similarities of the eigenenergy and wavefunction distributions between the $\Delta_{\mathrm{g}}=0$ case in Fig. \ref{gFig2}c,d and the small-size regime of the finite-$\Delta_{\mathrm{g}}$ case in Fig. \ref{gFig3}a,b. The small differences between the values of $|\beta_1|$ and $|\beta_2|$ shown in Fig. \ref{gFigWfc}b at small system sizes implies that $r_{1,2} = \sqrt{|\beta_2|/|\beta_1|}$ in Eq. \eqref{psiXsymNDg0a} is only slightly larger than 1. Therefore, at small values of $N < N_c$, the $r_{1,2}^{\pm (N+1)/2}$ terms in the boundary conditions $\psi(\pm (N+1)/2)=(0,0)^{\mathrm{T}}$ at the two ends of the system are approximately equal to 1 . Correspondingly,  the resultant eigenstates and eigenenergies  deviate very slightly from those in the $\Delta_{\mathrm{g}}=0$ case. However, as $N$ increases, the  $r_{1,2}^{\pm (N+1)/2}$ terms in the boundary conditions begin to have a significant impact. At the critical system size $N_{\mathrm{c}}$  the $r_{1,2}^{\pm (N_{\mathrm{c}+1)/2}}$ terms become large enough that it is no longer possible for  all four $|c_i|$s values to have approximately the same magnitude and $|\beta_1|$ be approximately equal to $|\beta_2|$ while satisfying the boundary conditions. The system is thus driven to transition to the large-size regime.

The size-dependent transition of the eigenenergy spectra in the finite $\Delta_{\mathrm{g}}$ case is in marked difference to the  $\Delta_{\mathrm{g}}=0$ case, in which the eigenenergies remain confined to the real energy axis regardless of the system size (Fig. \ref{gFig2}c). This difference can be attributed to the fact that  there are only two distinct eigenspinors in the $\Delta_{\mathrm{g}}=0$ case, and hence  a wavefunction that satisfies the boundary conditions can be  constructed out of only the two bulk eigenstates that share the same eigenspinor. By contrast,  in the finite $\Delta_{\mathrm{g}}$ case, since all four eigenspinors are distinct they would all contribute to the wavefunction to satisfy the boundary conditions at both edges. Additionally, the convergence of the eigenenergy spectrum to the GBZ in the thermodynamic limit can be explained by the following \cite{rafi2024twisted}: The wavefunction of an eigenstate satisfying the boundary conditions can generically be written as $\psi(x) = \sum_i c_i \beta_i^x (1, \chi_i)^\mathrm{T}$ where $c_i$ is the weight of the bulk eigenstate corresponding to $\beta_i$ where $|\beta_1|\leq|\beta_2|...|\beta_4|$ , and $(1,\chi_i)$ is the corresponding eigenspinor. 

As $x\rightarrow-\infty$ at the left edge of the system, $\beta_4^x$ becomes negligibly small compared to the other three terms and can be ignored. In order for the boundary condition $\psi(x \rightarrow -\infty)=(0,0)^{\mathrm{T}}$ at that edge to be satisfied, we require $c_j\beta_j^x$, $j\in(1,2,3)$ to have comparable magnitudes so that the three terms can cancel off one another. Likewise to fulfill the boundary condition at the right edge of the system as $\psi(x\rightarrow +\infty) = (0,0)^{\mathrm{T}}$,  we also require $a_l\beta_l^x$, $l \in (2,3,4)$ to have comparable magnitudes so that the three terms can cancel off one another. Considering both boundary conditions, the requirement that $c_j\beta_j^x$ for $j\in (2,3)$ have similar magnitudes in both the left and right edges leads to the equality $|\beta_2|=|\beta_3|$. This is because $c_3|\beta_3|^x$ ($c_2|\beta_2|^x$) will become overwhelmingly large otherwise compared to $c_2|\beta_2|^x$ ($c_3|\beta_3|^2$) as $x \rightarrow \infty$  ($x\rightarrow -\infty$). The eigenenergy spectrum of the system therefore tends towards the GBZ of the system, which is the locus of the equality $|\beta_2|=|\beta_3|$, as the system size increases. The convergence of $|\beta_2|$ and $|\beta_3|$ towards each other when $N$ exceeds $N_c$ is also evident from Fig. \ref{gFigWfc}b.   Notice that the argument above assumes that all four bulk eigenstates contribute to the OBC wavefunction that satisfies the boundary conditions. This assumption does not hold in the $\Delta_{\mathrm{g}}=0$ case where only two bulk eigenstates contribute as mentioned earlier.

\begin{figure*}[htp!]
  \centering
    \includegraphics[width=0.65\textwidth]{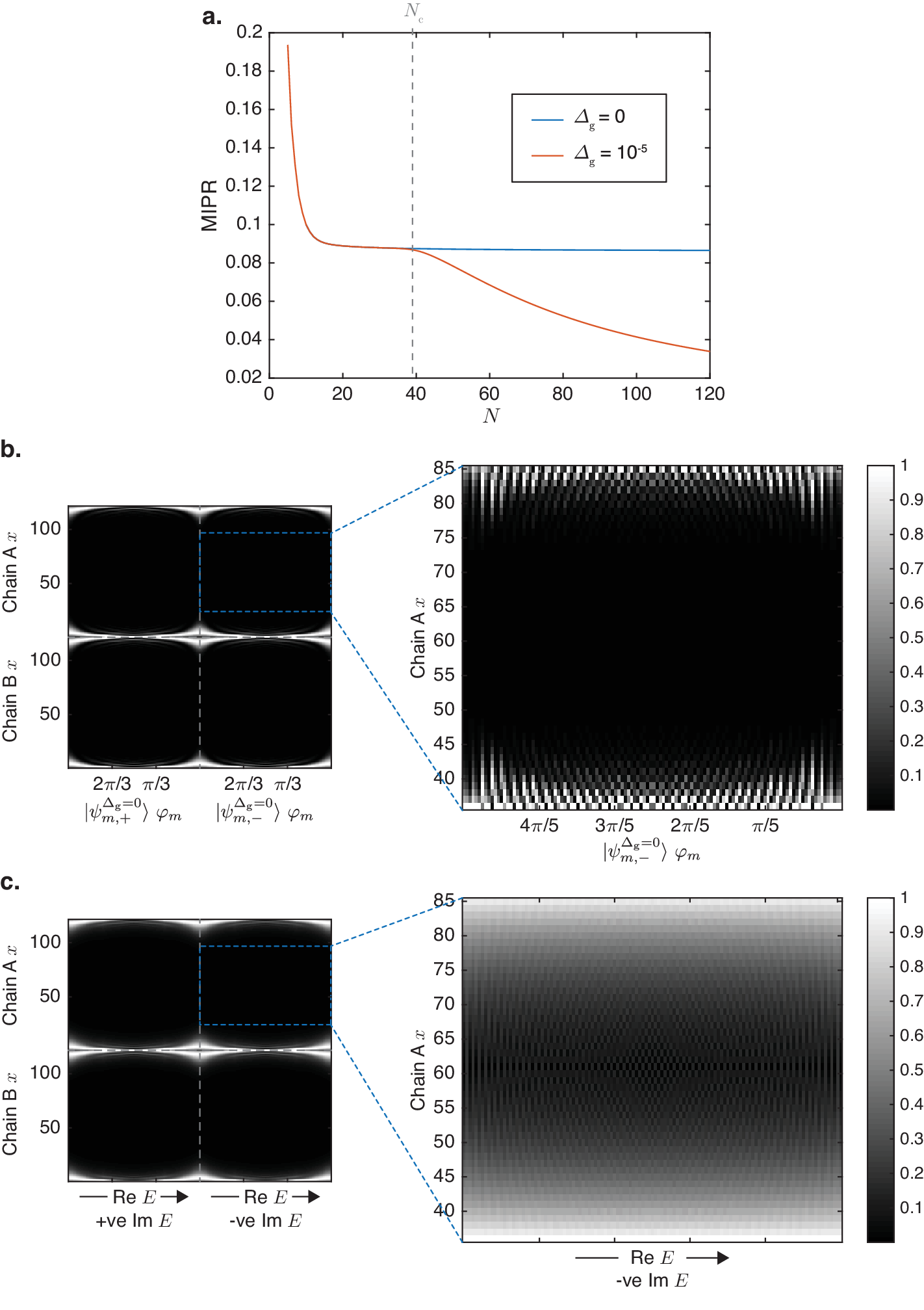}
  \caption{ (a) MIPR of cross-coupled chain systems with $\Delta_{\mathrm{g}}=10^{-5}$ and $\Delta_{\mathrm{g}}=0$, and $C_1=1.5$ and $C_{\mathrm{c}}=4.7$. (b) Eigenstate wavefunction amplitude for a system with $N=120$ and $\Delta_{\mathrm{g}}=0$. The right plot shows a magnified view of the wavefunction amplitude within the interior of Chain A in the region denoted by the dotted box. The wavefunctions are normalized so that the maximum amplitude within the region shown is 1. (c) Corresponding plot to b for a system with $\Delta_{\mathrm{g}}=10^{-5}$.    }
  \label{gFig4}
\end{figure*}

Another distinction between the $\Delta_{\mathrm{g}}=0$ and finite $\Delta_{\mathrm{g}}$ cases is that in the latter, the NHSE localization decreases with increasing system size above $N_{\mathrm{c}}$ whereas it remains approximately independent of the system size in the former. To explain this contrasting behavior, we introduce a quantitative measure of the NHSE localization  via the mean inverse participation ratio (MIPR) \cite{zhang2021non}, which is defined as
\begin{equation}
	\mathrm{MIPR} = \frac{1}{N} \sum_m  \left( \frac{ \sum_x |\psi_{m}(x)|^4}{ \left(\sum_x |\psi_m(x)|^2\right)^2} \right)
\end{equation}
where $\psi_m(x)$ is the value of the wavefunction of the $m$th OBC eigenstate at position $x$. A larger MIPR indicates a higher extent of NHSE localization. Fig. \ref{gFig4}a shows that below the critical size $N_{\mathrm{c}}$, the MIPRs for the finite and zero $\Delta_{\mathrm{g}}$ cases are very similar. However, when the system size exceeds $N_c$, the two MIPR plots diverge. The MIPR for the $\Delta_{\mathrm{g}}=0$ case tends to some finite asymptotic limit whereas for the finite $\Delta_{\mathrm{g}}$ case it continues to decrease beyond $N_{\mathrm{c}}$. In the latter case, the eigenenergy spectrum tends towards the complex energy plane GBZ, which coincides with the PBC spectrum. The PBC spectrum is, by definition, the loci of bulk eigenenergies with real values of $k$, which correspond to zero NHSE inverse localization length. This implies that as the system size increases beyond $N_c$, the inverse localization length of the wavefunction approaches to zero and thus the exponential decay of the wavefunction away from the edges becomes less prominent. This can be seen from a comparison of the right plots in Fig. \ref{gFig4}b ($\Delta_g = 0$ case) and \ref{gFig4}c (finite $\Delta_g$ case) for the $N=120$. (The axis labels in Fig. \ref{gFig4}b are plotted in order of decreasing $\varphi_m$ while those in Fig. \ref{gFig4}c are plotted in order of decreasing $\mathrm{Re}\ E$. This is because whereas $\varphi_m = m\pi/(N+1)$ with positive real integer $m$ is an appropriate quantum number for numbering the eigenstates in the $\Delta_g=0$ case, it is not a conserved quantum number in the finite $\Delta_g$ case in Fig. \ref{gFig4}c where the $\beta$ values are \textbf{not} given by $\exp(i\varphi_m)$ with real and integer $m$. For convenience we label the eigenstates by their eigenvalues instead. The eigenvalues in Fig. \ref{gFig4}b are plotted in decreasing values of $\varphi_m$ because for $0<\varphi_m<\pi/2$, $E_m \propto \cos(\varphi_m)$ increases with decreasing $\varphi_m$ so that each eigenstate in Fig. \ref{gFig4}c tends to the eigenstate in Fig. \ref{gFig4}b directly above it in the limit that $\Delta_g\rightarrow 0$.)

There is  very evident exponential decay of the wavefunctions from the edges for the $\Delta_{\mathrm{g}}=0$ case shown in Fig. \ref{gFig4}b while the exponential decay is reduced by a large extent in the finite $\Delta_{\mathrm{g}}$ case shown in Fig. \ref{gFig4}c. (There is still some localization near the boundaries in the left plots of Fig. \ref{gFig4}c because of the finite weights of the $\beta_1$ and $\beta_4$ states, which are localized at the left and right boundaries, respectively,  in $\psi(x)=\sum_i |\chi_i\rangle c_i \beta_i^x$. )

\section{Conclusion}
In summary, we have investigated the coupled system consisting of two reciprocal 1D chains with a non-reciprocal cross inter-chain coupling. One of the most salient features of the system is  the exponential localization of the eigenstates near the edges of the chains akin to the NHSE even when each of the individual uncoupled chains are themselves Hermitian. Interestingly, the eigenenergy distribution of the system without any gain/loss terms deviates from the complex energy plane GBZ in the thermodynamic-limit, unlike that of conventional 1D non-Hermitian systems. However, the introduction of even an infinitesimal amount of gain / loss to the system would cause the thermodynamic-limit eigenenergy distribution to revert to the conventional behaviour and coincide with the GBZ. Another interesting feature of the system in the presence of finite gain/loss is a critical phenomenon in which the eigenenergy distribution exhibits a abrupt switchover beyond a critical system size from  the real energy axis to an elliptical profile approaching the PBC spectrum. 

We provided the analytical explanation underlying this critical phenomenon. Finally, we found that the degree of NHSE localization, as characterized by the system's MIPR ratio, is correlated to its eigenenergy distribution. In the zero-$\Delta_{\mathrm{g}}$ case where the eigenenergy distribution is confined to the real energy axis, the NHSE is highly localized and remains so regardless of the system size. While in the finite-$\Delta_{\mathrm{g}}$ case, the degree of NHSE localization drops sharply when the system size exceeds the critical value and the eigenenergy distribution approaches the PBC spectrum. Our study provides new insights into critical phenomena related to the eigenenergy and skin mode localization in coupled 1D systems, especially the role of the inter-chain coupling, which has hitherto not been investigated in depth. The phenomena can be practically realized in the TE circuit platform presented here as well as in other platforms used in the study of non-Hermitian systems such as photonics, optics, metamaterials, and acoustics systems.

\subsubsection*{Acknowledgements}
This work is supported by the Ministry of Education (MOE) Tier-II Grant MOE-T2EP50121-0014 (NUS Grant No. A-8000086-01-00), and MOE Tier-I FRC Grant (NUS Grant No. A-8000195-01-00).

\section*{Appendix: Do all type inter-chain coupling  induce NHSE? } 
\begin{figure*}[htp!]
  \centering
    \includegraphics[width=0.6\textwidth]{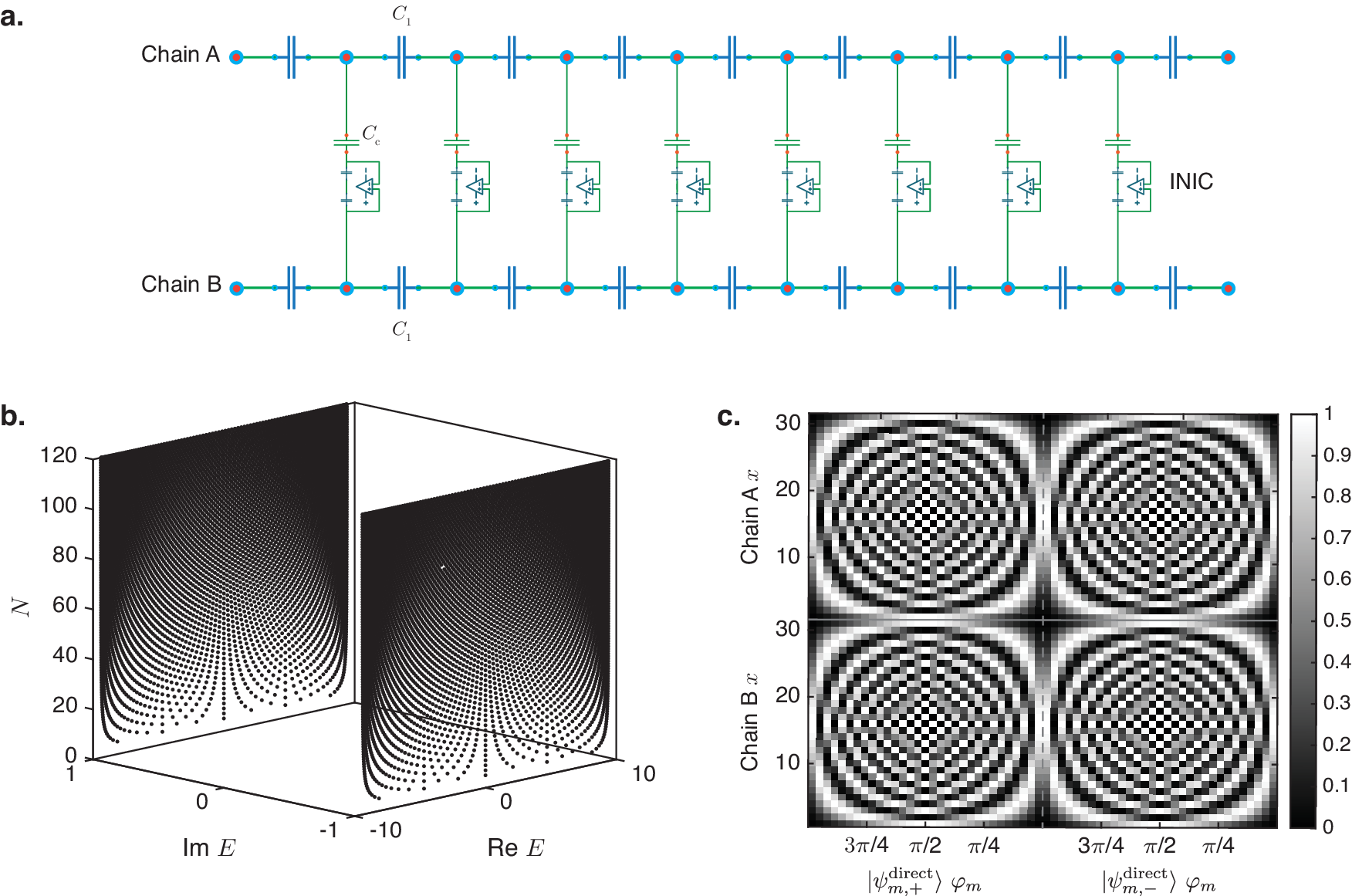}
  \caption{ (a) Schematic representation of a pair of 1D Hermitian chains with a non-reciprocal inter-chain coupling that does not exhibit the NHSE. The uncoupled chains have reciprocal couplings of $C_1$ between each node and its left and right neighbors on the same chain. The INIC realizes a non-reciprocal direct coupling between each node in a chain and its counterpart on the other chain with a coupling magnitude of $C_\mathrm{c}$ and a phase difference of $\pi$ between the coupling of chain A to chain B, and that of chain B to chain A. (b) Eigenspectrum of system in a. as a function of the system size $N$ with $C_1=4.7$ and $C_{\mathrm{c}}=1.5$. (c) Wavefunction amplitude of the system in c. for $N=31$. }
  \label{gFig5}
\end{figure*}

In this section, we provide an example to show that not all types of non-reciprocal inter-chain coupling induces NHSE in a coupled system of Hermitian subsystems. We consider a system of two Hermitian 1D chains  in which the two chains are directly coupled to each other (i.e., each node in one chain is coupled to its counterpart on the other chain) via a non-reciprocal inter-chain coupling $\pm C_{\mathrm{c}}$  (see Fig. \ref{gFig5}a). The corresponding bulk Hamiltonian of this system is given by 
\begin{equation}
H^{\mathrm{direct}}(\beta)= \begin{pmatrix}
C_1 \left( \beta + \frac{1}{\beta}\right) & C_c 
\\ -C_c & C_1 \left( \beta + \frac{1}{\beta}\right) \end{pmatrix}
\label{appeq1}
\end{equation}

The solutions of $\beta$ for a given eigenenergy $E$ are given by 
\begin{equation}
  \beta^{\mathrm{direct}}_{s_c,s_d} = \frac{E - i s_c C_{\mathrm{c}} +  s_d \sqrt{  (E-i s_c C_{\mathrm{c}})^2 - 4C_1^2}}{2C_1} \label{betaEqDir}
\end{equation} 
where $s_c$ and $s_d$ independently take the values of $\pm 1$, and the eigenspinors $|\chi^{\mathrm{direct}}_{s_c,s_d}\rangle$ associated with $\beta^{\mathrm{direct}}_{s_c,s_d}$ are independent of $E$ and given by 
\begin{equation}
  |\chi^{\mathrm{direct}}_{s_c,s_d}\rangle = \begin{pmatrix} s_ci  \\ 1 \end{pmatrix}. \label{chiEqDir}
\end{equation}

From Eq. \eqref{betaEqDir}, it is evident that when $E - i s_c C_{\mathrm{c}}$ is real and $|E-i s_c C_{\mathrm{c}}|^2 < 2|C_1|$, the $\beta^{\mathrm{direct}}_{s_c, \pm 1}$s form a complex conjugate pair where
\begin{align}
  |\beta_{s_c, \pm 1}| &= 1, \\
  \mathrm{arg} \left(\beta_{s_c, s_d}\right) &= s_d \varphi^{\mathrm{direct}}_{s_c},\ \varphi^{\mathrm{direct}}_{s,c} \equiv \cos^{-1}  \left(\frac{E - i s_c C_{\mathrm{c}}}{2C_1}\right).
\end{align}

The wavefunctions for a pair of degenerate eigenstates $\psi^{\mathrm{direct}}_{m, s_c}(x)$ with the eigenenergies $E^{\mathrm{direct}}_{m, s_c}$ that satisfies the boundary conditions  $\psi^{\mathrm{direct}}_{m, s_c}(0)=\psi^{\mathrm{direct}}_{m, s_c}(N+1)=(0,0)^{\mathrm{T}}$ can thus be obtained as 
\begin{align} 
  \psi^{\mathrm{direct}}_{m, s_c}(x) &= \sin(\varphi_m x) \begin{pmatrix}s_ci \\ 1 \end{pmatrix} \label{psiDirect} \\
  E^{\mathrm{direct}}_{m, s_c} &= 2C_1\cos(\varphi_m) + i s_c C_{\mathrm{c}}
\end{align}
where $\varphi_m = m\pi/(N+1)$.

The OBC eigenspectrum of Eq. \eqref{appeq1} hence takes the form of two parallel lines on the complex energy plane with the imaginary parts of $\pm C_1$ and real parts between $\pm 2C_1$ (Fig. \ref{gFig5}b). In contrast to the cross-coupled system with the eigenstate wavefunctions given by Eq. \eqref{psiDg0} where the cross-coupling induces a NHSE in the originally Hermitian uncoupled chains, there is no NHSE in the wavefunctions Eq. \eqref{psiDirect} here because the constituent $\beta$ components in the wavefunction all have unit moduli (Fig. \ref{gFig5}c). This example shows that not all forms of non-reciprocal inter-chain coupling between Hermitian chains will result in NHSE. 


%

\end{document}